\documentclass[12pt]{article}

\usepackage{amsfonts}
\usepackage{amsmath}
\usepackage{epsfig}
\usepackage{latexsym}
\usepackage{graphicx}
\usepackage{subfigure}
\usepackage{amssymb}
\numberwithin{equation}{section}
\allowdisplaybreaks

\def\spa#1{\phantom{\fbox{\rule[-#1cm]{0cm}{0cm}}}}

\def\be{\begin{equation}}
\def\ee{\end{equation}}
\def\bea{\begin{eqnarray}}
\def\eea{\end{eqnarray}}

\def\half{{1\over 2}}

\def\del{\partial}
\def\nn{\nonumber}

\renewcommand{\thefootnote}{\fnsymbol{footnote}}

\begin{document}

\hfuzz=100pt
\title{{\Large \bf{$n$-DBI gravity}}}
\date{}
\author{Carlos Herdeiro$^a$\footnote{herdeiro@ua.pt}, Shinji Hirano$^b$\footnote{hirano@eken.phys.nagoya-u.ac.jp} and Yuki Sato$^{b,c}$\footnote{ysato@th.phys.nagoya-u.ac.jp}
  \spa{0.5} \\
\\
$^a${\small{\it Departamento de F\'isica da Universidade de Aveiro and I3N}}
\\ {\small{\it Campus de Santiago, 3810-183 Aveiro, Portugal}}\\
\\
$^{b}${\small{\it Department of Physics, Nagoya University}}
%\\ {\small{\it Nagoya University}}
\\ {\small{\it Nagoya 464-8602, Japan}}\\
\\
$^{c}${\small{\it The Niels Bohr Institute}}
\\ {\small{\it Blegdamsvej 17, DK-2100, Copenhagen, Denmark}}
%\\ {\small{\it Copenhagen, Denmark}}
  \spa{0.3} 
}
\date{}

\maketitle
\centerline{}

\begin{abstract}
$n$-DBI gravity is a gravitational theory introduced in \cite{Herdeiro:2011km}, motivated by Dirac-Born-Infeld type conformal scalar theory and designed to yield non-eternal inflation spontaneously. It contains a foliation structure provided by an everywhere time-like vector field ${\bf n}$, which couples to the gravitational sector of the theory, but decouples in the small curvature limit. We show that any solution of Einstein gravity with a particular curvature property is a solution of $n$-DBI gravity. Amongst them is a class of geometries isometric to Reissner-Nordstr\"om-(Anti) de Sitter black hole, which is obtained within the spherically symmetric solutions of $n$-DBI gravity minimally coupled to the Maxwell field. These solutions have, however, two distinct features from their Einstein gravity counterparts: 1) the cosmological constant appears as an integration constant and can be positive, negative or vanishing, making it a \textit{variable} quantity of the theory; 2) there is a non-uniqueness of solutions with the same total mass, charge and effective cosmological constant. Such inequivalent solutions cannot be mapped to each other by a foliation preserving diffeomorphism. Physically they are distinguished by the expansion and shear of the congruence tangent to ${\bf n}$, which define scalar invariants on each leave of the foliation. 
\end{abstract}

\renewcommand{\thefootnote}{\arabic{footnote}}
\setcounter{footnote}{0}

\newpage

%%%%%%%%%%%%%%%%%%%%%%%%%%%%%%%%%%%%%%%%%
\section{Introduction}
The mystery surrounding dark energy and dark matter, as well as the expected detection of gravitational radiation in the near future, and consequent observation of strong gravity dynamics, are powerful motivations to explore gravitational theories beyond Einstein's general relativity. In the most optimistic scenarios some distinct phenomenological signature of Einstein's gravity or of some alternatives/extension thereof could be identified. Moreover, such studies, commonly shed light on general relativity itself. An example is the extension of general relativity to higher dimensions, from which one understands how special the uniqueness theorems for black hole solutions of general relativity are. One other example is higher curvature gravity, from which one understands how special the fact that the equations of motion of general relativity are only second order in derivatives is.

In this paper we further explore an extension of general relativity recently proposed by some of us  \cite{Herdeiro:2011km}. This gravitational theory was motivated by conformal invariance and the realisation that a Dirac-Born-Infeld (DBI) type conformal scalar theory has a phenomenologically interesting dynamics when its degree of freedom is interpreted as the conformal factor of a conformally flat universe. Such cosmological model can yield non-eternal (primordial) inflation spontaneously, smoothly connected to a radiation and matter era and a subsequent (present) accelerated expansion. The two distinct accelerating periods, with two distinct effective cosmological constants, are a manifestation that the cosmological constant can vary in this theory. Moreover, a large hierarchy between these two cosmological constants can be naturally achieved, if the naive cosmological constant appearing in a weak curvature expansion of the theory is associated to the TeV scale, suggesting also a new mechanism to address the cosmological constant problem.

 The extension of this model to include all gravitational field degrees of freedom (rather than just an overall conformal factor) suggested the introduction of an everywhere time-like unit vector field ${\bf n}$, coupled to the gravitational sector (but not necessarily to the matter sector) of the theory. We therefore dub this theory `$n$-DBI gravity'. Such coupling leads to the breakdown of Lorentz invariance, but since ${\bf n}$ decouples from the gravitational dynamics in the weak curvature limit, Lorentz invariance is restored and Einstein gravity is recovered in this limit. This breakdown of Lorentz invariance at strong curvature and restoring at weak curvature is somewhat reminiscent of what was proposed in Horava-Lifschitz gravity \cite{Horava:2009uw}; in our theory, however, it is explicit.

Mathematically, the introduction of ${\bf n}$, allows $n$-DBI gravity - which has an infinite power series in the Ricci curvature - to yield equations of motion which are \textit{at most} second order in time, albeit higher order in spatial derivatives. This is in principle a desirable property, so as to avoid ghosts in the quantum theory, a property normally associated to only Lovelock gravity \cite{Lovelock}, of which Einstein gravity is a particular example. Thus, our model, illustrates one other path to achieve this property. Also, the existence of this time-like vector field naturally induces a foliation of space-time. The allowed diffeomorphisms of the theory should preserve ${\bf n}$ and hence this foliation, which will therefore be only a subgroup of general coordinate transformations. Physically this means that there are quantities,  defined on each leave, which have no invariant meaning in Einstein gravity but become invariant under this symmetry group; two that shall be discussed are the shearing and expansion of the congruence tangent to ${\bf n}$.

Remarkably, we shall show that many solutions of Einstein's gravity are also solutions of $n$-DBI gravity, despite the fact that the latter has an infinite power series in scalar curvature. Moreover, this is not only the case for vacuum solutions, or even Einstein spaces. We shall actually explicitly show that the Reissner-Nordstr\"om black hole with (or without) a cosmological constant solves the field equations of $n$-DBI gravity. In our theory, however, this is a larger family of solutions than in Einstein gravity, not only parameterised by total mass, charge and cosmological constant, but also by the \textit{slicing}. This is a novel type of non-uniqueness, which, albeit trivial from the geometric viewpoint, has physical significance in this framework. Also, in agreement with the aforementioned, the cosmological constant for these solutions will arise as an integration constant and can therefore take different values.

This paper is organised as follows. Section 2 defines $n$-DBI gravity, introducing further details as compared to those presented in \cite{Herdeiro:2011km}. Section 3 addresses spherically symmetric solutions, keeping some technicalities to Appendix A. We draw conclusions in Section 4.

%%%%%%%%%%%%%%%%%%%%%%%%%%%%%%%%%%%%%%%%%
\section{Formulation of $n$-DBI gravity}\label{model}

$n$-DBI gravity \cite{Herdeiro:2011km} is defined by the action
\begin{align}
S=-{1\over 16\pi G_N}\int d^4x\left\{\frac{12\lambda}{G_N}\sqrt{-g}\biggl[\sqrt{1+{G_N\over 6\lambda}\left(\mbox{}^{\mbox{}^{(4)}\!}\! R+{\cal K}\right)}-q\biggr]+16\pi G_N\mathcal{L}_{\rm matter}\right\}\ ,\label{DBIgravity}
\end{align}
where $G_N$ is Newton's constant, $\mbox{}^{\mbox{}^{(4)}\!}\! R$ is the four dimensional Ricci scalar constructed in the standard way from the space-time metric $g_{\mu\nu}$ and  $\mathcal{L}_{\rm matter}$ is the matter Lagrangian density. The theory includes two dimensionless constants: $\lambda$ and $q$. We have introduced a scalar quantity closely related to the Gibbons-Hawking-York boundary term \cite{Gibbons:1976ue, York:1972sj}:
\be
{\cal K}:=-{2\over \sqrt{h}}n^{\sigma}\del_{\sigma}\left(\sqrt{h}K\right)
=-2\left(K^2+n^{\sigma}\del_{\sigma}K\right)\ .
\ee
This quantity is specified by the choice of a unit time-like vector ${\bf n}$, and the consequent first and second fundamental forms (induced metric and extrinsic curvature)
\begin{align}
h_{\mu\nu}\equiv g_{\mu\nu}+n_{\mu}n_{\nu}\ ,\qquad\qquad 
K_{\mu\nu}\equiv \half n^{\sigma}\del_{\sigma}h_{\mu\nu}\ ,
\end{align}
from which $K=K_{\mu\nu}h^{\mu\nu}$, the trace of the extrinsic curvature, is constructed. 

We assume the time-like vector ${\bf n}$ defines a foliation of space-time, $\mathcal{F}$, by space-like leaves. Being hyper-surface orthogonal, it is natural to adopt the ADM decomposition of space-time \cite{Arnowitt:1960es}
\be
ds^2=-N^2 dt^2+h_{ij}\left(dx^i+N^i dt\right)\left(dx^j+N^j dt\right)\ .\label{ADM}
\ee
Standard nomenclature is that $N$ and $N^i$ are the \textit{lapse} and \textit{shift}. Then, as a 1-form, ${\bf n}=-Ndt$ and therefore ${\bf n}\wedge d{\bf n}=0$, in agreement with Frobenius' theorem. In terms of the ADM decomposition, the normal derivative is expressed as
\be
n^{\sigma}\del_{\sigma}=N^{-1}\left(\del_t-\pounds_N\right) \ ,
\ee
using the Lie derivative $\pounds_N$ for the vector field $N^i$; also, the extrinsic curvature reads 
\be
K_{ij}={1\over 2N}\left(\dot{h}_{ij}-\nabla_iN_j-\nabla_jN_i\right)\ .
\ee
Moreover, to reexpress the theory in terms of the ADM variables $N$, $N^i$ and $h_{ij}$ note that  \cite{York:2004gb}: 
%\be
%\mbox{}^{(4)}\! R=\left(R+K_{ij}K^{ij}-K^2\right)+{2\over\sqrt{h}}N^{-1}\del_0\left(\sqrt{h}K\right)-{2\over\sqrt{h}}N^{-1}\del_i
%\left(\sqrt{h}\left(N^iK+\nabla^iN\right)\right)\ ,
%\ee
%
%
\be
\mbox{}^{\mbox{}^{(4)}\!}\! R+{\cal K}=R+K_{ij}K^{ij}-K^2-2N^{-1}\Delta N\equiv \mathcal{R}\ , \label{defr}
\ee
where $R$ is the Ricci scalar of $h_{ij}$. 
%Therefore
%and the action is reexpressed as
%
%
%\begin{align}
%S=-{1\over 16\pi G_N}\int d^4x\left\{ \frac{12\lambda}{G_N}\sqrt{h}N\left[\sqrt{1+{G_N\over 6\lambda}
%\mathcal{R}}-q\right]+16\pi G_N\mathcal{L}_{\rm matter}\right\}\ .\label{DBIADM}
%\end{align}
Then, varying the action with respect to $N$, $N^i$ and $h_{ij}$ one obtains, respectively, the Hamiltonian constraint
\begin{align}
\hspace{-.3cm}
{1+{G_N\over 6\lambda}\left(R-N^{-1}\Delta N\right)\over 
\sqrt{1+{G_N\over 6\lambda}{\cal R}}}-q
-\frac{G_N}{6\lambda}\Delta\left(1+{G_N\over 6\lambda}{\cal R}\right)^{-1/2}=-\frac{4\pi G_N^2}{3\lambda \sqrt{h}}\frac{\delta \mathcal{L}_{\rm matter}}{\delta N} \ ,\label{Nvariation}
\end{align}
the momentum constraints
\be
\nabla^j\left({K_{ij}-h_{ij}K \over \sqrt{1+{G_N\over 6\lambda}{\cal R}}}\right)=-\frac{8\pi G_N}{\sqrt{h}}\frac{\delta \mathcal{L}_{\rm matter}}{\delta N^i} \ ,
\ee
and the $h_{ij}$ evolution equation
%\begin{align}
%\hspace{-.3cm}
%(-\partial_t+N^l\nabla_l)\left({K^{ij}-\gamma^{ij}K \over \sqrt{1+{G_N\over 6\lambda}{\cal R}}}\right)-2\left({K^{l(i}-\gamma^{l(i}K \over \sqrt{1+{G_N\over 6\lambda}{\cal R}}}\right)\nabla_lN^{j)}\\
%+\frac{N(-R^{ij}+KK^{ij}-2K^{il}K^j_{\ l}+\nabla^i\nabla^jN+\gamma^{ij}(K_{mn}K^{mn}-N^{-1}\Delta N))}{ \sqrt{1+{G_N\over 6\lambda}{\cal R}}}\\
%+(N\nabla^i\nabla^j-\gamma^{ij}(\nabla_l N)\nabla^l)\left(1+{G_N\over 6\lambda}{\cal R}\right)^{-1/2}=0 \ ;
%\end{align}
%The latter equation may be written, more elegantly as
\begin{align}
\hspace{-.3cm}
\frac{1}{N}(\partial_t-\pounds_N)\left({K^{ij}-h^{ij}K \over \sqrt{1+{G_N\over 6\lambda}{\cal R}}}\right)=(\nabla^i\nabla^j-h^{ij}(\nabla_l \ln N)\nabla^l)\left(1+{G_N\over 6\lambda}{\cal R}\right)^{-1/2}\nn \\
+\frac{-R^{ij}+KK^{ij}-2K^{il}K^j_{\ l}+N^{-1}\nabla^i\nabla^jN+h^{ij}(K_{mn}K^{mn}-N^{-1}\Delta N)}{ \sqrt{1+{G_N\over 6\lambda}{\cal R}}}+\frac{16\pi G_N}{N\sqrt{h}}\frac{\delta \mathcal{L}_{\rm matter}}{\delta h_{ij}} \ . \label{gammavariation}
\end{align}

In $n$-DBI gravity, the space-time manifold is a differentiable manifold $\mathcal{M}$, with an additional structure: an everywhere time-like vector field ${\bf n}$, defining a co-dimension one foliation $\mathcal{F}$, which should be regarded as defining part of the topology of $\mathcal{M}$. The leaves of $\mathcal{F}$ are hypersurfaces of constant time. The allowed diffeomorphisms in $n$-DBI gravity must preserve $\mathcal{F}$. They are dubbed \textit{foliation preserving diffeomorphisms} (FPDs) \cite{Horava:2009uw} and are denoted by Diff$_{\mathcal{F}}(\mathcal{M})$. In the ADM coordinates, FPDs are generated by
\be
t\rightarrow t +\xi^0(t) \ , \qquad x^i\rightarrow x^i+\xi^i(t,x^j) \ . 
\ee
It follows that under FPDs the ADM fields transform as
\begin{align}
\delta N & =   N\dot{\xi}^0+\xi^i\partial_i N+\xi^0\dot{N} \ , \nn \\ 
\delta N_i & = N_j\partial_i\xi^j+\xi^j\partial_jN_i+\dot{\xi}^jh_{ij}+\dot{\xi}^0N_i+\xi^0\dot{N}_i \ , \nn \\
\delta h_{ij} & =\xi^0\dot{h}_{ij}+\xi^k\partial_k h_{ij}+h_{ki}\partial_j\xi^k+h_{kj}\partial_i\xi^k \ .
\end{align}

The second fundamental form $K_{ij}$ transforms covariantly under Diff$_{\mathcal{F}}(\mathcal{M})$. Consequently, scalar quantities such as $K$ or $K_{ij}K^{ij}$ will be non-vanishing in all allowed frames if non-vanishing in a particular frame. A convenient way to think about these invariants is to regard the traceless $K_{ij}$ and its trace K as the shear, $\sigma_{ij}$, and expansion, $\theta$, of the congruence of time-like curves with tangent ${\bf n}$ (there is no rotation since the ${\bf n}$ is hypersurface orthogonal):
\be
\sigma_{ij}=K_{ij}-\frac{1}{3}Kh_{ij} \ , \qquad \theta=K \ . \ee

To exemplify the loss of symmetry, as compared to Einstein gravity, consider the \textit{non}-FPD 
\be 
t\rightarrow \tilde{t}=t+f(r) \ , \label{nonfpd}
\ee
acting on Minkowski space-time in spherical coordinates: $ds^2=-dt^2+dr^2+r^2d\Omega_2$ (which is of course a solution of vacuum $n$-DBI gravity). If one  requires a vanishing trace for the extrinsic curvature of the resulting foliation, $K=0$, the resulting line element becomes completely defined up to an integration constant $C_3$:
\begin{align}
ds^2= & -\left(1+\frac{C_3}{r^4}\right)dt^2 +\left(\frac{dr}{\sqrt{1+\frac{C_3}{r^4}}}+\sqrt{\frac{C_3}{r^4}}dt\right)^2+r^2d\Omega_2 \ . \label{c3}
\end{align}
The fact that the constant $C_3$ cannot be gauged away in $n$-DBI gravity (i.e that (\ref{nonfpd}) is a prohibited transformation) is now manifest in the fact that  $K_{ij}K^{ij}=\sigma_{ij}\sigma^{ij}=6C_3/r^6$ is non-vanishing. Thus, for $C_3\neq 0$, $r=0$ is a \textit{shearing singularity}, which is a physical singularity, since the shear cannot be gauged away by transformations of Diff$_{\mathcal{F}}(\mathcal{M})$. As we shall see below this foliation is actually a solution of vacuum $n$-DBI gravity, which is not the case for the foliation obtained by (\ref{nonfpd}) with a generic $f(r)$. In other words: non-FPDs generically map solutions to geometrically equivalent space-times which are nevertheless {\it not} solutions of the theory. In special cases, however, as illustrated here, a non-FPD maps a solution to another solution, albeit physically inequivalent. In spirit, such non-FPDs, are analogous to an electromagnetic duality transformation, or more general duality transformations found, for instance, in string theory.

\subsection{Einstein gravity limit}

Taking the Einstein gravity limit, $\lambda\rightarrow \infty$ and $q\to 1$ with $\lambda(1-q)$ fixed, the action (\ref{DBIgravity}) becomes
\begin{align}
S=-{1\over 16\pi G_N}\int_{\mathcal{M}} d^4x\left\{\sqrt{-g}\biggl[\mbox{}^{\mbox{}^{(4)}\!}\! R-2G_N\Lambda_{\rm Einstein}\biggr]+16\pi G_N\mathcal{L}_{\rm matter}\right\}+\frac{1}{8\pi G_N}\int_{\partial M} d^3x \sqrt{h} K\ ,\label{Einsteingravity}
\end{align}
where the effective cosmological constant is
\be
\Lambda_{\rm Einstein}= \frac{6\lambda(q-1)}{G_N^2} \ , \ee
and the Gibbons-Hawking-York term is taken in a hypersurface orthogonal to ${\bf n}$. Equivalently, 
 (\ref{Nvariation})-(\ref{gammavariation}) reduce to the corresponding equations of Einstein gravity with a   cosmological constant (see eg. \cite{Gourgoulhon:2007ue}):
\be
R+K^2-K_{ij}K^{ij}=2G_N\Lambda_{\rm Einstein}
%-\frac{16\pi G_N}{\sqrt{h}}\frac{\delta \mathcal{L}_{\rm matter}}{\delta N}
 \ ,\label{E1}\ee
\be
\nabla^j\left({K_{ij}-h_{ij}K }\right)=0 \ ,
\ee
\be
\frac{1}{N}(\partial_t-\pounds_N)K_{ij}=N^{-1}\nabla_i\nabla_j N-R_{ij}-KK_{ij}+2K_{il}K_j^{\ l} \ , \label{E3}
\ee
where we have neglected the matter terms, for simplicity. As can be seen in both the action and in the equations of motion, in this limit, there is no coupling between the gravitational dynamics and ${\bf n}$. 
Thus, covariance under the {\it full} diffeomorphisms ${\rm Diff}({\cal M})$ is restored and so is Lorentz invariance. Even without taking the limit, whenever the curvature ${\cal R}$ is very small, the equations (\ref{Nvariation})-(\ref{gammavariation}) are well approximated by these Einstein equations. This means that Lorentz invariance is restored to a very good accuracy in weakly curved space-times of $n$-DBI gravity.

\subsection{Solutions with constant $\mathcal{R}$}
Taking $\mathcal{R}$ to be constant, and dubbing $\sqrt{1+G_N\mathcal{R}/(6\lambda)}\equiv C$, the equations of motion  (\ref{Nvariation})-(\ref{gammavariation}) reduce to
\begin{align}
\hspace{-.3cm}
R-N^{-1}\Delta N+\frac{6\lambda}{G_N}(1-qC)=-\frac{8\pi G_NC}{ \sqrt{h}}\frac{\delta \mathcal{L}_{\rm matter}}{\delta N} \ ,\label{NvariationR}
\end{align}
\be
\nabla^j\left({K_{ij}-h_{ij}K }\right)=- \frac{8\pi G_NC}{\sqrt{h}}\frac{\delta \mathcal{L}_{\rm matter}}{\delta N^i} \ ,
\ee
\begin{align}
\hspace{-.3cm}
\frac{1}{N}(\partial_t-\pounds_N)\left({K^{ij}-h^{ij}K}\right)=h^{ij}(K_{mn}K^{mn}-N^{-1}\Delta N)\nn \\
{-R^{ij}+KK^{ij}-2K^{il}K^j_{\ l}+N^{-1}\nabla^i\nabla^jN}+\frac{16\pi G_NC}{N\sqrt{h}}\frac{\delta \mathcal{L}_{\rm matter}}{\delta h_{ij}} \ . \label{gammavariationR}
\end{align}
The momentum constraints and the dynamical equations are equivalent to those of Einstein gravity, but with a renormalisation of the matter terms by a factor of $C$. The Hamiltonian constraint is also equivalent to that of Einstein gravity with, besides the renormalisation by $C$ of the matter term, a cosmological constant
\be
\Lambda_{C}=\frac{3\lambda}{G_N^2}(2qC-1-C^2) \ . \label{lambdac}
\ee
As discussed in section 3, this cosmological constant can be positive, negative or zero, depending on the choices of $C$ (and the value of $q$). 

We are thus led to the following theorem and corollary:

\bigskip

{\bf Theorem:} \textit{Any solution of Einstein gravity with a cosmological constant plus some matter Lagrangian admitting a foliation with constant $\mathcal{R}$, as defined in (\ref{defr}), is a solution of $n$-DBI gravity (with an appropriate renormalisation of the solution parameters).}

\bigskip

{\bf Corollary:} \textit{Any Einstein space (hence solution of the Einstein equations with a cosmological constant) admitting a foliation such that $R-N^{-1}\Delta N$ is constant - where $R$ and $\Delta$  are the Ricci scalar and the Laplacian of the 3-metric $h_{ij}$ and $N$ is the lapse in the ADM decomposition -  is a solution of $n$-DBI gravity (with an appropriate renormalisation of the solution parameters).}

%%%%%%%%%%%%%%%%%%%%%%%%%%%%%%%%%%%%%%%%%
\section{Spherically symmetric solutions of $n$-DBI gravity}\label{solutions}

The most generic spherically symmetric line element reads
\be
ds^2=-g_{TT}(R,T)dT^2+g_{RR}(R,T)dR^2+2g_{TR}(T,R)dTdR+g_{\theta\theta}(R,T)d\Omega_2 \ . 
\ee
Defining a new radial coordinate $r^2=g_{\theta\theta}(R,T)$ and a new time coordinate $t=t(R,T)$ it is possible to transform this line element into a standard diagonal form, with only two unknown functions: $g_{tt}(t,r)$ and $g_{rr}(t,r)$. Then, the vacuum Einstein equations yield, as the only solution, the Schwarzschild black hole, and as a corollary Birkhoff's theorem, namely that spherical symmetry implies staticity. In $n$-DBI gravity, however, only FPDs are allowed. Thus, only $t=t(T)$ is allowed. As a consequence, the most general line element compatible with spherical symmetry is:
\be
ds^2=-N^2(t,r)dt^2+e^{2f(t,r)}\left(dr+e^{2g(t,r)}dt\right)^2+r^2d\Omega_2 \ . \label{ansatz}
\ee
Herein we consider the case with only $r$ dependence for the three metric functions. To include the possibility of charge, we take
\be
\mathcal{L}_{\rm matter}= -\frac{1}{16\pi}\sqrt{-g}F_{\mu\nu}F^{\mu\nu} \ , \ee
where ${\bf F}=d{\bf A}$ is the Maxwell 2-form, and to address the purely electric case we take the ansatz:
\be
{\bf A}=A(r)dt \qquad \Rightarrow \qquad \mathcal{L}_{\rm matter}= \frac{r^2\sin\theta e^{-f}}{8\pi N}(A')^2\ .\ee

To directly solve the equations of motion (\ref{Nvariation})-(\ref{gammavariation}) is quite tedious. It proves more convenient to consider the reduced system obtained by specialising the action (\ref{DBIgravity}) to our ansatz. One obtains the effective Lagrangian
\be
\mathcal{L}_{eff}=\lambda r^2e^{f}N\left[\sqrt{1+\frac{G_N}{6\lambda}\mathcal{R}}-q\right] +\frac{G_N^2}{6N}r^2e^{-f}(A')^2\ , \label{efelag}
\ee
whose equations of motion are a subset of the full set of constraints (\ref{Nvariation})-(\ref{gammavariation}).\footnote{One should check that the final solution satisfies the equations of motion (\ref{Nvariation})-(\ref{gammavariation}), which is indeed the case.} These equations of motion can be solved with full generality (see Appendix A), but it turns out that the most interesting solutions are the subset with $\mathcal{R}$ constant. These are given by

\begin{align}
ds^2= & -\left(1-\frac{2G_N M_1}{r}+{CQ^2\over  r^2}+\frac{C_3}{r^4}-\frac{\Lambda_1 r^2}{3}\right)dt^2\nn \\
& +\left(\frac{dr}{\sqrt{1-\frac{2G_NM_1}{r}+{CQ^2\over  r^2}+\frac{C_3}{r^4}-\frac{\Lambda_1 r^2}{3}}}+\sqrt{\frac{2G_NM_2}{r}+\frac{C_3}{r^4}+\frac{\Lambda_2r^2}{3}}dt\right)^2+r^2d\Omega_2 \ , \label{dRNdS}
\end{align}
where
\be
\Lambda_1\equiv \frac{2\lambda}{G_N}\left({qC}-1\right) \ , \qquad
\Lambda_2 \equiv \frac{\lambda}{G_N}\left({4qC}-1-3C^2\right) \ ,\ee
%with
%\begin{align}
%e^{2f}=&\left[1+{2M_1\over r}+{CQ^2\over  r^2}+{C_3\over r^4}-\frac{\Lambda_1 r^2}{3}
%\right]^{-1}\ ,\label{solhrr}\\
%e^{4g}=&e^{-2f}\left[{2M_2\over r}+{C_3\over r^4}+\frac{\Lambda_2 r^2}{3}
%\right]\ ,\label{solNr}
%\end{align}
and $Q,CM_1,M_2,C_3$ are integration constants. Moreover, as expected,
\be 
 \Lambda_1+\Lambda_2=\Lambda_C \ ,
\ee
where $\Lambda_C$ is defined in (\ref{lambdac}). This family of solutions is therefore characterised by these five integration constants and the two dimensional constants of the theory $(\lambda,q)$.

\subsection{Analysis of the solutions}

\subsubsection*{Isometry to Reissner-Nordstr\"om-(A)dS}
Performing a coordinate transformation $t\rightarrow T=T(t,r)$,
\be
dT=dt-\frac{1}{1-\frac{2G_NM}{r}+{CQ^2\over  r^2}-\frac{\Lambda r^2}{3}}\sqrt{\frac{\frac{2G_NM_2}{r}+\frac{C_3}{r^4}+\frac{\Lambda_2r^2}{3}}{1-\frac{2G_NM_1}{r}+{CQ^2\over  r^2}+\frac{C_3}{r^4}-\frac{\Lambda_1 r^2}{3}}}dr \ , \label{ct}
\ee
where $M\equiv M_1+M_2$, the line element (\ref{dRNdS}) becomes recognisable as the Reissner-Nordstr\"om-(Anti)-de-Sitter solution with mass $M$, charge $\sqrt{C}Q$ and cosmological constant $\Lambda_C$:
\begin{align}
ds^2=  -\left(1-\frac{2G_NM}{r}+{CQ^2\over  r^2}-\frac{\Lambda_C r^2}{3}\right)dT^2 +\frac{dr^2}{1-\frac{2G_NM}{r}+{CQ^2\over  r^2}-\frac{\Lambda_C r^2}{3}}+r^2d\Omega_2 \ . \label{RNdS}
\end{align}
Observe the renormalisation of the charge, as anticipated in Section 2. 

Geometrically, the solution we have found is nothing but this standard solution of Einstein gravity, written in an unusual set of coordinates that can be thought of as a superposition of Gullstrand-Painlev\'e and Schwarzschild coordinates. The coordinate transformation (\ref{ct}) is not, however, a foliation preserving diffeomorphism. Thus, in $n$-DBI gravity, (\ref{dRNdS}) and (\ref{RNdS}) describe the same solution if and only if $M_2=C_3=\Lambda_2=0$. Otherwise they are two distinct solutions with different physical invariants (discussed below) which {\it happen to} be mapped by a non-FPD.

\medskip
Notice that there is, for example, no Reissner-Nordstr\"om (A)dS solution in Gullstrand-Painlev\'e coordinates. If the symmetry group of $n$-DBI gravity was the set of general coordinate transformations, we would have found such solution. In other words, the breakdown of the symmetry to FPDs is explicitly reflected in the form of the solutions (\ref{dRNdS}). 

\subsubsection*{Expansion and shear}
As discussed above the shear and expansion transform covariantly under Diff$_{\mathcal{F}}(\mathcal{M})$. For the solution (\ref{dRNdS}) the scalar invariants read
\be
\theta =-\frac{3 \left(G_NM_2+\Lambda_2r^3 \right)}{\sqrt{C_3+2 G_NM_2 r^3+ \Lambda_2r^6}}\ ,\qquad
\sigma_{ij}\sigma^{ij}=6\left[\frac{C_3+G_NM_2 r^3}{r^3 \sqrt{C_3+2 G_NM_2 r^3+\Lambda_2r^6}}\right]^2 \ ;
%\sigma_{ij}=-\frac{C_3+M_2 r^3}{r^3 \sqrt{C_3+2 M_2 r^3+\Lambda_2r^6}}{\rm diag}
%\{-2h_{rr}, h_{\theta\theta}, h_{\varphi\varphi}\}\ .
\ee
we also have
\be
K_{ij}K^{ij}-K^2=6\left(\frac{C_3}{r^6}-\Lambda_2\right)\ .
\ee
It is manifest that $M_2$, $\Lambda_2$ and $C_3$ enter in these scalar invariants and therefore have physical meaning in defining the solution. As usual in Einstein gravity, one can invoke smoothness to rule out some solutions as unphysical. For instance smoothness of the constant (Riemann) curvature spaces $(M_1=M_2=Q=0)$, requires $C_3=0$ to avoid the shearing singularity at $r=0$.

\subsubsection*{Asymptotic behaviour}
Asymptotically ($r\rightarrow \infty$) the solution (\ref{dRNdS}) becomes a constant curvature space
\be
R_{\mu\nu\alpha \beta}=\frac{\Lambda_C}{3}\left(g_{\mu\alpha}g_{\nu\beta}-g_{\mu\beta}g_{\nu\alpha}\right)  \ .  \label{cc}
\ee
With appropriate choices of $C$ one may set either $\Lambda_1=0$ or $\Lambda_2=0$, keeping the other non-vanishing. In both cases one recognises de Sitter space: either written in Painlev\'e-Gullstrand coordinates (with cosmological constant $\Lambda_2$), or written in static coordinates (with cosmological constant $\Lambda_1$). In the latter case, Anti-de-Sitter space may also occur, written in global coordinates. Keeping both $\Lambda_1$ and $\Lambda_2$ non-vanishing one has an unusual slicing of constant curvature spaces.  This can represent de Sitter space-time, Anti-de Sitter space-time or Minkowski space, depending on the sign of the total cosmological constant $\Lambda_1+\Lambda_2$. Indeed, the integration constant $C$ controls the magnitude of the cosmological constant: 
\begin{align}
C  & \ \in \ \ ]q-\sqrt{q^2-1}, q+\sqrt{q^2-1}[ \ , \ \ \ & {\rm de \ Sitter}\nn \\
C & \ = q\pm \sqrt{q^2-1} \ ,  & {\rm Minkowski} \nn \\
C &  \ {\rm otherwise}  & {\rm AdS} \ . \nn
\end{align}
dS and Minkowski space solutions can only exist if $q\ge 1$.

\section{Conclusions and Discussion}\label{conclusion}
In this paper we have explored some further properties of $n$-DBI gravity, beyond those studied in \cite{Herdeiro:2011km}, which focused on cosmology. 

A crucial property of the theory is the existence of an everywhere time-like vector field ${\bf n}$. We have assumed it to be hyper-surface orthogonal - which is expressed by the relation we have chosen between ${\bf n}$ and the ADM quantities - albeit this condition could be dropped and an even more general framework considered. The existence and role played by ${\bf n}$  is reminiscent of Einstein-Aether theory (see \cite{Jacobson:2008aj} for a review). 

It follows that the symmetry group of the theory is that which preserves ${\bf n}$ and therefore the foliation defined by it, $\mathcal{F}$. This group, denoted by  Diff$_{\mathcal{F}}(\mathcal{M})$, is the group of FPDs, and it is therefore smaller than general coordinate transformations; it is the group that leaves invariant the equations of motion. This means that a non-FPD  applied to a solution of $n$-DBI gravity is \textit{not}, in general, a solution of $n$-DBI gravity. Exceptionally, however, this may happen and a non-FPD may map two solutions. These should be regarded, however, as physically distinct solutions, perhaps in the same orbit of a larger symmetry group, in the same spirit of many duality symmetries or solution generating techniques that have been considered in the context of supergravity or string theory. In $n$-DBI gravity it is unclear, at the moment, if such larger symmetry group exists, but an explicit example of a non-FPD mapping (inequivalent) solutions was provided by (\ref{ct}). The solutions are, of course, isometric; in this particular example they are the standard Reissner-Nordstr\"om-(A)dS geometry in two different coordinate systems. Observe, however, the non-trivial dynamics of the theory, where the mass and the cosmological constant can in effect be split between two slicings but not the charge.

The fact that the spherically symmetric solutions of $n$-DBI gravity minimally coupled to a Maxwell field contain precisely the Reissner-Nordstr\"om geometry (with or without a cosmological constant) is remarkable and, as far as we are aware, unparalleled, within theories of gravity with higher curvature terms. This leads to the natural question of how generic is it that Einstein gravity solutions are solutions of $n$-DBI gravity (with the same matter content)? Following the theorem and corollary presented in Section 2 this question can be recast very objectively as the existence of a foliation with a specific property. How generically can such foliation be found? Can it be found for the Kerr solution?

Finally, as discussed at the beginning of Section 3, the ansatz compatible with spherical symmetry in $n$-DBI gravity has more degrees of freedom than in Einstein gravity. It will be quite interesting to see if, even in vacuum, such ansatz can accommodate a non-trivial time dependence, prohibited in Einstein gravity by Birkhoff's theorem, which would manifest the existence of a scalar graviton mode.

%%%%%%%%%%%%%%%%%%%%%%%%%%%%%%%%%%%%%%%%%%%%%%%
\section*{Acknowledgement}

S. H. would like to thank Masaki Shigemori for useful conversations. Y. S. would like to thank the Niels Bohr Institute for their warm hospitality. 
This work was partially supported by the Grant-in-Aid for Nagoya
University Global COE Program (G07) and by FCT (Portugal) through the project CERN/FP/116341/2010.

%%%%%%%%%%%%%%%%%%%%%%%%%%%%%%%%%%%%%%%%%%%%%%%%

\appendix
\renewcommand{\theequation}{\Alph{section}.\arabic{equation}}
\section{Derivation of the solutions}
Taking the ansatz
\be
ds^2=-N^2(r)dt^2+e^{2f(r)}\left(dr+e^{2g(r)}dt\right)^2+r^2d\Omega_2 \ , \label{ansatzr}
\ee
it follows that ($'\equiv d/dr$):
\be
K_{ij}=-\frac{e^{2g}}{N} \, {\rm diag}\left\{e^{2f}(f+2g)',r,r\sin^2\theta\right\} \ , \qquad K=-\frac{e^{2g}}{N}\left(\frac{2}{r}+(f+2g)'\right) \ ,
\ee
\be
R_{ij}= {\rm diag}\left\{\frac{2f'}{r},\frac{rf'+e^{2f}-1}{e^{2f}}, \sin^2\theta \frac{rf'+e^{2f}-1}{e^{2f}}\right\} \ , \qquad R=\frac{2e^{-2f}}{r^2}\left(2rf'+e^{2f}-1\right) \ ,
\ee
Thus, 
\be
\mathcal{R}=\frac{2}{r^2}\left[1-(re^{-2f})'-\frac{(re^{4g})'}{N^2}-\frac{2rf'e^{4g}}{N^2}-\frac{e^{-f}}{N}(r^2N'e^{-f})'\right] \ . \label{curlyR}
\ee
From the effective Lagrangian (\ref{efelag}), the $A$ equation of motion can be solved straightaway to yield
\be 
A'=Q\frac{Ne^{f}}{r^2} \ , \label{aeq}
\ee
where $Q$ is an integration constant.  The $g$ equation of motion yields the compact relation
\begin{align}
&\left({1\over\sqrt{1+\frac{G_N}{6\lambda}\mathcal{R}}}\right)'={(f+\ln N)'\over\sqrt{1+\frac{G_N}{6\lambda}\mathcal{R}}}\ , \label{eq2}
\end{align}
and can be integrated to yield
\be 
\left(1+\frac{G_N}{6\lambda}\mathcal{R}\right)^{-1/2}=\frac{Ne^{f}}{C} \ , \label{eq2v2}\ee
where $C$ is an integration constant. More explicitly, this equation may be written as
\be
\left(re^{4g+2f}\right)'=N^2e^{2f}\left(1-(re^{-2f})'\right)-Ne^{f}\left(r^2e^{-f}N'\right)'
+{3\lambda\over G_N}r^2\left(N^2e^{2f}-C^2\right) \ . \label{geq}
\ee
The $f$ and $N$ equations of motion, upon using (\ref{aeq}) and (\ref{eq2v2}), read, respectively
\begin{align}
\left(re^{4g+2f}\right)'=&re^{-2f}\left(e^{2f}N^2\right)'-{Ne^{f}\over 2}\left(r^2e^{-f}N'\right)'+{r^2e^{-f}N'\over 2}(Ne^f)'\nn\\
&-{3\lambda\over G_N}r^2\left(C^2-CqNe^{f}\right)+{G_NCQ^2Ne^{f}\over 2 r^2}\ ,\label{feq}\\
\left(r e^{4g+2f}\right)'=&-{1\over 4}\left(r^2\left(e^{-2f}\right)'N^2e^{2f}\right)'-{3\lambda\over G_N}r^2\left(C^2-CqNe^{f}\right)+{G_NCQ^2Ne^{f}\over 2r^2}\ . \label{Ham}
\end{align}
Eq. (\ref{Ham}) is the Hamiltonian constraint (\ref{Nvariation}), after using (\ref{eq2v2}) and (\ref{geq}).

\subsection{Solutions with constant $\mathcal{R}$}
To proceed we take the combination $Ne^f=\tilde{C}={\rm  constant}$ which implies that $\mathcal{R}$ is constant. We shall address the general solution in the next subsection, but it turns out that the most interesting solution are found in this subset. With this choice, we observe from eq. (\ref{eq2v2}) that $\mathcal{R}={\rm constant}$. From the resulting equations of motion, equating (\ref{geq}) with either (\ref{feq}) or (\ref{Ham}) (which become identical), we find the ODE:
\be
Y''+{6\over r}Y'+{4\over r^2}Y={12\lambda\over G_N}\left(1-{qC\over \tilde{C}}\right)
-{2G_NCQ^2\over \tilde{C}r^4}\ ,
\ee
where $Y\equiv e^{-2f}-1$. It is now straightforward to obtain the exact solution. It reads (\ref{dRNdS}), where $\tilde{C}$ has been eliminated by rescaling $C$ and the time coordinate.

\subsection{Generic solution}

Since the set of equations we are solving is a second order ODE with three unknowns, we expect a total of six integration constants. The constant ${\cal R}$ solution exhibited below has only five integration constants and thus it is not the most general one. The latter can be obtained observing that the equations (\ref{feq}) and (\ref{Ham}) imply
\be
-\left(r^{-2}(\log N)'\right)'=\left(r^{-2}f'\right)'\qquad\quad\Longrightarrow\qquad\quad
Ne^f=\tilde{C}e^{{1\over 3}C_4 r^3} \ .\label{Nef}
\ee
$C_4$ is the sixth integration constant, which was absent in the constant ${\cal R}$ solution. 
Similarly to the constant ${\cal R}$ case, it is straightforward to find a second order ODE:
\be
W''+{6\over r}W'+{4\over r^2}W={4e^{2C_4r^3\over 3}\over r^2}+{12\lambda\over G_N}\left(e^{2C_4r^3\over 3}-{qCe^{C_4r^3\over 3}\over \tilde{C}}\right)-{2G_NCQ^2e^{C_4r^3\over 3}\over \tilde{C}r^4}\ ,
\ee
where we defined $W\equiv e^{-2\left(f-{C_4r^3\over 3}\right)}$. 
This can be integrated to give explicit solutions. As they are not very illuminating, however, we will not present them here. Indeed, the solutions with $C_4\ne 0$ seem rather exotic, since (\ref{eq2v2}) and (\ref{Nef}) imply that their asymptotic behavior at $r=+\infty$ is very different from that of Einstein gravity: the $C_4<0$ solutions have a curvature singularity at $r=+\infty$ and thus we regard these solutions as unphysical; the $C_4>0$ solutions have the maximal negative curvature ${\cal R}=-6\lambda/G_N$ at $r=+\infty$. Although they are interesting in their own right, we shall not discuss these solutions further herein.

%%%%%%%%%%%%%%%%%%%%%%%%%%%%%%%%%%%%%%%%%%%%%%

\end{document}